# A solution to the inverse problem of non-coherent image propagation

**Jan A. Mamczur [1*], Marek J. Matczak [2]**

[1] *Department of Physics, Rzeszow University of Technology, ul. W. Pola 2, 35-959 Rzeszow, Poland*
[2] *Institute of Physics, University of Rzeszow, ul. Rejtana 16A, 35-310 Rzeszow, Poland*
*\*Corresponding author: janand@prz.edu.pl*

**Abstract:** We derive a formula for the non-coherent wave field propagation in terms of the Fourier transformation. As a result, we find a theoretical solution of the inverse problem of image propagation in non-coherent case. However, the practical solution of the inverse problem in a numerical way leads to rapidly decreasing SNR. We also discuss origin and behavior of that phenomenon. Some numerical examples are presented.

**1. Introduction**

The inverse problem of non-coherent image propagation is meant in this paper as determining the original non-coherent wave field intensity distribution in a plane on the basis of knowledge of intensity distribution of the wave field arising from the original one as a result of its propagation at a distance $z$ in the free space. It means that in such reconstruction process one can only use intensity distribution $I(x,y)$ of the image propagated instead of the quick-varying complex amplitude $a(x,y,t)$ of non-coherent wave field in the propagated-image plane. In the latter case, the original non-coherent image may be recovered by a lens. However if one knows the intensity $I(x,y)$ only, e.g. an image mapped without any lens by a photosensitive matrix placed a distance $z$ away from the original image, then no lens system will recover the original non-coherent image.

The inverse problem of coherent image propagation has been solved in the algebraic way by W. D. Montgomery [1]. A solution to the coherent inverse problem in terms of self-imaging phenomenon has been proposed in [2,3]. The solution to the non-coherent inverse problem, discussed in this paper, results from the formula for non-coherent wave field propagation. The formula is derived in Section 2, and the discussed solution is presented in Section 3.

## 2. The non-coherent wave field propagation operator

In case of non-coherent wave field propagation, a linear optical system shown in Fig. 1 can be described by the linear transformation of wave field intensity distribution [4]:

$$I(x, y; z) = \hat{P}_z I(x, y; 0) = \int\int_{-\infty}^{+\infty} g(x-x', y-y'; z) I(x', y'; 0) \, dx' dy', \qquad (1)$$

where $I(x',y';0)$ and $I(x,y;z)$ are the wave field intensity distributions in the plane $z=0$ and in the plane $z>0$, respectively, and $\hat{P}_z$ is the propagation operator.

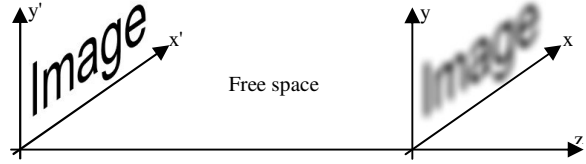

Fig. 1. A sketch of the geometry of the optical system.

The integral operator kernel $g(x,y;z)$ can be expressed as follows [4]:

$$g(x, y; z) = \kappa |h(x, y; z)|^2, \qquad (2)$$

where $\kappa$ is a real constant, and $h(x,y;z)$ is the kernel of the propagation operator acting on the wave field complex amplitude $a(x,y)$ in the coherent case:

$$a(x, y; z) = \int\int_{-\infty}^{+\infty} h(x-x', y-y'; z) a(x', y'; 0) \, dx' dy' \qquad (3)$$

In the non-coherent case, two wave field analytic signals $u(x,y,z,t)$ taken at different points of a plane are statistically independent [4]:

$$\langle u(x_1, y_1, t) \, u^*(x_2, y_2, t) \rangle = \kappa I(x_1, y_1) \delta(x_1 - x_2, y_1 - y_2), \qquad (4)$$

where the sign $\langle \, \rangle$ denotes averaging over time, and $\delta$ is the Dirac delta function.

In order to get a diagonal form of the non-coherent wave field propagation operator, we can apply to Eq. (1) the two-dimensional Fourier transformation and the convolution theorem, which yields:

$$J(\omega_x, \omega_y; z) = G(\omega_x, \omega_y; z) J(\omega_x, \omega_y; 0), \qquad (5)$$

where $J(\omega_x,\omega_y;0)$ and $J(\omega_x,\omega_y;z)$ are Fourier transforms of the non-coherent wave field intensity distributions $I(x,y;0)$ and $I(x,y;z)$, respectively, $G(\omega_x,\omega_y;z)$ is the Fourier transform of the kernel $g(x,y;z)$ multiplied by $2\pi$, and $\omega_x$, $\omega_y$ are spatial angular frequencies having the sense of wave vector projections on the axes $x$ and $y$, respectively.

In case of the coherent wave field propagation at the distance $z$ in the free space, the propagation-operator kernel $h(x,y;z)$ is determined by the Helmholtz-Kirchhoff diffraction integral [5]:

$$h(x, y; z) = \frac{1}{2\pi} \frac{\exp\left(ik\sqrt{x^2 + y^2 + z^2}\right)}{\sqrt{x^2 + y^2 + z^2}} \left[ ik - \frac{1}{\sqrt{x^2 + y^2 + z^2}} \right] \frac{z}{\sqrt{x^2 + y^2 + z^2}}. \qquad (6)$$

Using Eq. (2), we obtain:

$$g(x, y; z) = \frac{\kappa k^2}{4\pi^2} \frac{z^2}{\left(x^2 + y^2 + z^2\right)^2} + \frac{\kappa}{4\pi^2} \frac{z^2}{\left(x^2 + y^2 + z^2\right)^3}, \tag{7}$$

where $k$ is the wave number.

In order to find two-dimensional Fourier transform of the kernel $g(x,y;z)$, we shall take advantage of the rotational symmetry of $g$. Inserting new variables $r,\varphi$ in spatial domain and $\rho,\psi$ in spatial angular frequency domain:

$$x + iy = re^{i\varphi}, \qquad \omega_x + i\omega_y = \rho e^{i\psi}, \tag{8}$$

and assuming that a given function $f(x,y)$ fulfills the rotational symmetry condition, i.e. $f(x,y) = \phi(r)$, one can determine its Fourier transform as the Hankel transform (of the zero-th order) [6]. Using Hankel transformation, we have found Fourier transform of the propagation operator kernel (7):

$$2\pi \mathcal{F}[g(x,y;z)] = \frac{\kappa k^2}{4\pi} z\rho K_1(z\rho) + \frac{\kappa}{16\pi} \rho^2 K_2(z\rho) \stackrel{\text{def}}{=} G(\rho;z), \tag{9}$$

where $K_1$ and $K_2$ are the modified Bessel functions of the second kind (MacDonald functions) of the first and second order, respectively.

Equation (5) together with the transform $G(\rho;z)$ is a more convenient tool for numerical calculations than Eq. (1) with the kernel given by Eq. (7), because it gives diagonal form of the non-coherent propagation operator and it enables us to take advantage of FFT (Fast Fourier Transform) efficiency.

Starting from the following properties [7] of the Bessel functions discussed:

$$\lim_{\rho \to 0} z\rho K_1(z\rho) = 1, \quad \lim_{\rho \to 0} z^2 \rho^2 K_2(z\rho) = 2, \tag{10}$$

and requiring $J(0,0;z) = J(0,0;0)$, i.e. conservation of the wave field total power during propagation, which is equivalent to $G(0; z) = 1$, we get the explicit form of the coefficient $\kappa$:

$$\kappa = \frac{4\pi}{k^2}\left(1 + \frac{1}{2k^2 z^2}\right)^{-1}. \tag{11}$$

Thus, the final form of the spatial-spectrum transfer function $G(\rho;z)$ is as follows:

$$G(\rho;z) = \left(1 + \frac{1}{2k^2 z^2}\right)^{-1}\left[z\rho K_1(z\rho) + \frac{\rho^2}{4k^2} K_2(z\rho)\right]. \tag{12}$$

Figure 2 presents plots of $G(\rho;z)$ as functions of spatial frequency $\rho$ for different propagation distances $z$. The spatial frequency is expressed by the multiple of the wave number $\rho' = \rho/k$, likewise the propagation distance is expressed by the multiple of the wavelength $z' = z/\lambda$. The plots illustrate a rapid suppression of higher spatial frequencies by the spatial-spectrum transfer function $G(\rho';z')$.

The function $G(\rho;z)$ given by Eq. (9) consists of two summands. They correspond, with conservation of order, to both the summands of the non-coherent propagation kernel given by Eq. (7). However, for the remote field, i.e. when the propagation distance $z$ is much greater then the wavelength $\lambda$, we can reduce Eq. (7) as well as Eq. (9) to their first summands.

For large argument values $z\rho \gg 1$, it is worth using the approximation [7] of the modified Bessel function, which yields:

$$G(\rho;z) \approx \frac{\kappa k^2}{4\sqrt{2\pi}}\left(1 + \frac{\rho}{4zk^2}\right)\sqrt{z\rho}\, e^{-z\rho}\left[1 + \Theta\left(\frac{1}{z\rho}\right)\right]. \tag{13}$$

where $\Theta\left(\frac{1}{x}\right)$ denotes a little magnitude of the same order as $\frac{1}{x}$.

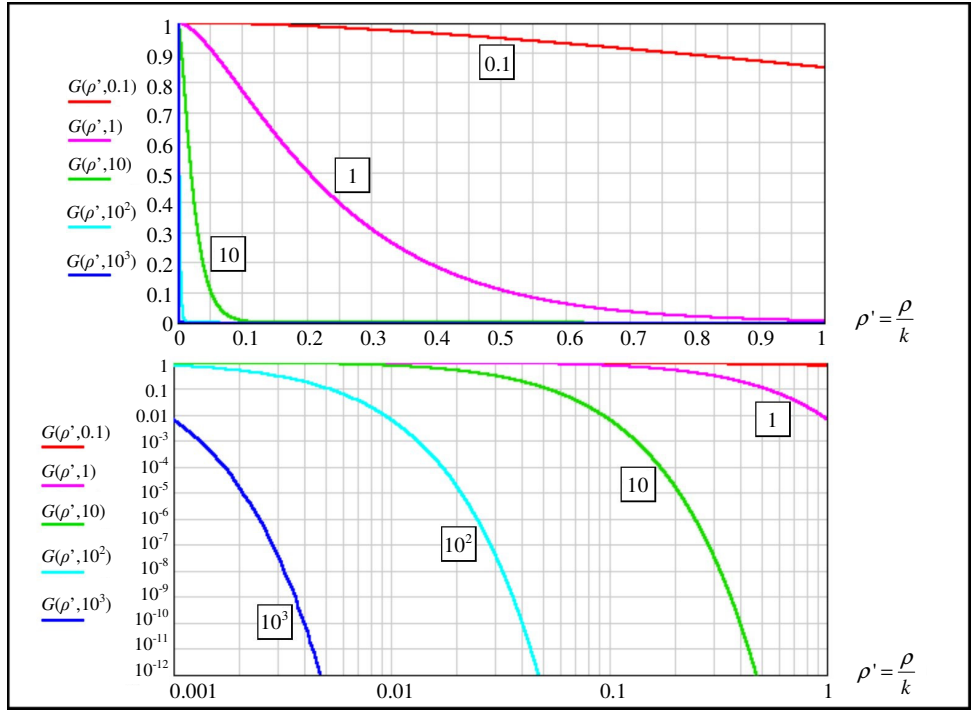

Fig. 2. $G(z',\rho')$ as a function of $\rho'=\rho/k$ for different $z'=z/\lambda$, in linear and logarithmic scale.

## 3. Inverse problem of non-coherent image propagation

As it has been shown in the previous section, the spatial-spectrum transfer function $G(\rho;z)$ is a continuous function and it takes only positive values. It leads to the existence of an inverse propagation operator, action of which can be expressed in Fourier representation in the following form:

$$\tilde{J}(\omega_x,\omega_y;0)=\left[G(\omega_x,\omega_y;z)\right]^{-1}J(\omega_x,\omega_y;z)\ ,\qquad(14)$$

which results directly from Eq. (5). The inverse Fourier transform $\tilde{I}(x,y;0)$ of $\tilde{J}(\omega_x,\omega_y;0)$ should give the original wave field intensity distribution $I(x,y;0)$ from before propagation.

In order to verify usability of such an operator, a practical solution to the inverse problem for a series of propagation distances has been realized numerically by means of Eq. (14) and Fourier transformations. In all analyzed cases, the wave field intensity distribution $I(x,y;z)$ resulting from propagation was computed on the basis of Eq. (1) applied to the original distribution $I(x,y;0)$. Practical realization of the image-reconstruction procedure requires recording the propagated image $I(x,y;z)$ into a computer using analog-to-digital conversion (ADC):

$$I(x,y;z)\ \rightarrow\ I_D(x,y;z)=I(x,y;z)+N(x,y)\qquad(15)$$

where $I_D$ denotes the digital form of the propagated-image intensity distribution and $N$ is the quantization noise [8]. Computational result $\tilde{J}_D$ of the reconstruction described by Eq. (14) takes the following form:

$$\tilde{J}_D(\omega_x,\omega_y;0) = \left[G(\omega_x,\omega_y;z)\right]^{-1}\left[J(\omega_x,\omega_y;z)+J_N(\omega_x,\omega_y)\right]$$
$$= \tilde{J}(\omega_x,\omega_y;0)+\tilde{J}_N(\omega_x,\omega_y) \qquad (16)$$

where $J_N$ denotes the quantization-noise Fourier transform.

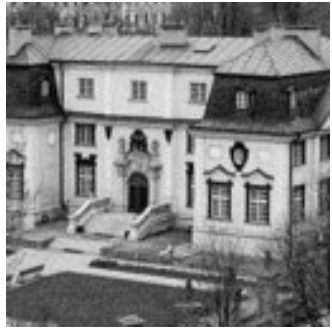

Fig. 3. The original image, $I(x,y;0)$.

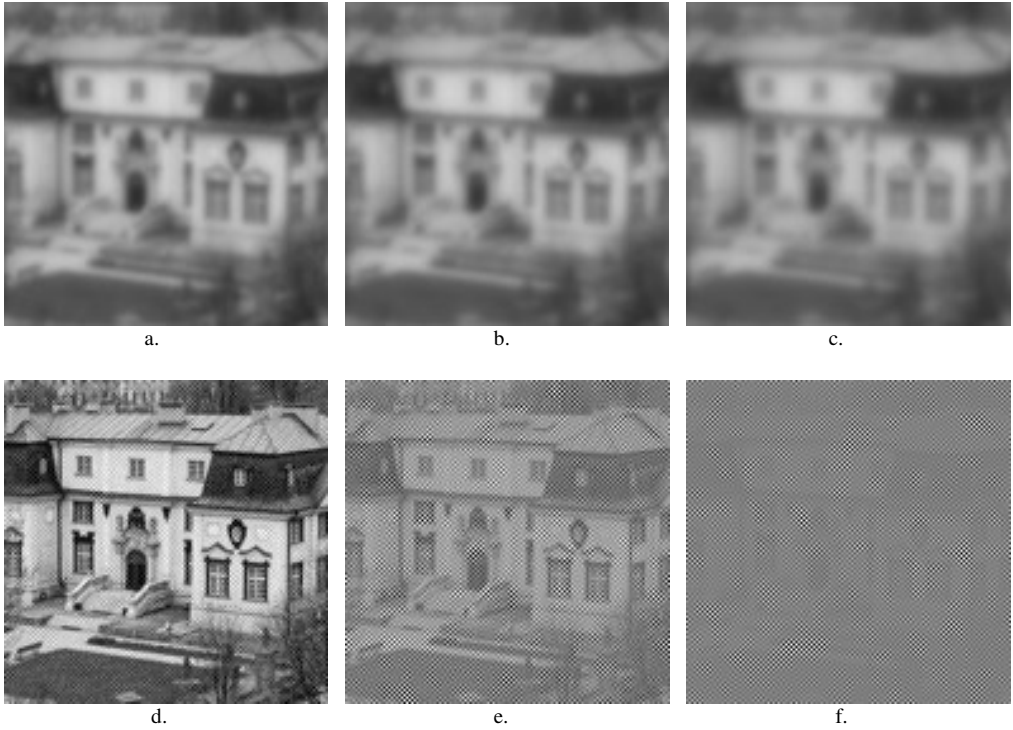

Fig. 4. Examples of image reconstruction for three different propagation distances close to the distance $z_o$; a., b. and c.: images $I(x,y;z)$ after propagation at $z = 0.79\,z_o$, $z = z_o$ and $z = 1.19\,z_o$, respectively; d. (SNR=9.7), e. (SNR=1.0) and f. (SNR=0.11): images $\tilde{I}_D(x,y;0)$ reconstructed from the images shown in a., b. and c., respectively. Absolute size of the image $LxL = 10\text{x}10$ mm$^2$, the wavelength $\lambda=0.5$ μm, $z_o = 0.225$ mm; digitalization: resolution 128x128 pixels, 12-bit quantization of ADC.

Unfortunately, effectiveness of the image reconstruction by means of the inverse propagation operator described above is limited to a certain propagation distance $z_o$ at which the signal-to-noise ratio (SNR) in the reconstructed image is equal to unity in the linear scale or equal to zero (dB) in the logarithmic one. In surrounding of the distance $z_o$, SNR in the reconstructed image rapidly decreases. The distance $z_o$ depends on the number of bits of ADC and on the greatest spatial frequency $\rho_{max}$ in the original-image spatial spectrum. For $z \geq z_o$, a larger number of bits of ADC leads to a better reconstruction while increasing image spatial frequency $\rho_{max}$ (by increasing image resolution) gives a reverse effect.

Examples of image reconstruction for three different propagation distances close to the distance $z_o$ are presented in Fig. 4. The original image is shown in Fig. 3. SNR has been determined as the ratio of the image-intensity standard deviation to the noise standard deviation. The noise $N(x,y)$ has been obtained by subtracting the propagated-image intensity distribution $I(x,y;z)$ before ADC from that after ADC $I_D(x,y;z)$.

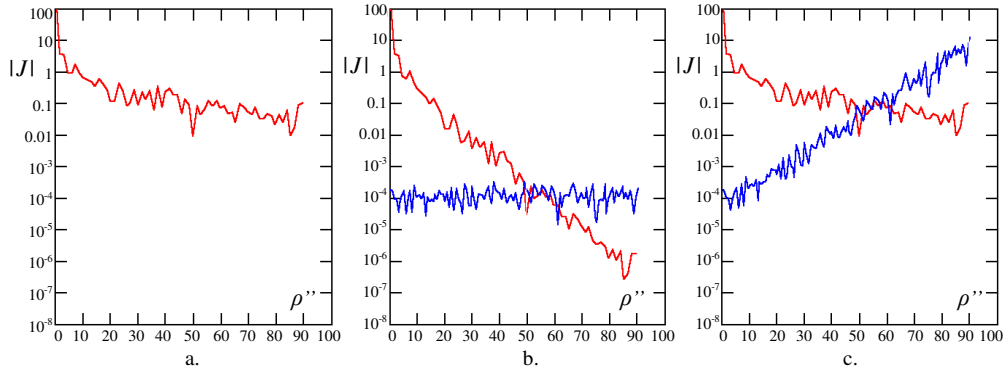

Fig. 5. Spatial spectrum of image intensity (red) and quantization noise (blue) in the diagonal cross-section of the spectrum plane for: original image in Fig. 3 (a. $J(\omega_x,\omega_y;0)$), propagated image in Fig. 4b and quantization noise (b. $J(\omega_x,\omega_y;z_0) + J_N(\omega_x,\omega_y)$), and reconstructed image with amplified quantization noise in Fig. 4e (c. $\tilde{J}(\omega_x,\omega_y;0) + \tilde{J}_N(\omega_x,\omega_y)$); $\rho'' = \rho/\rho_o$, $\rho_o = 2\pi/L = 628.3$ m$^{-1}$.

The results presented above and illustrated in Fig. 4 and 5 suggest that the lack of image reconstruction ability for higher propagation distances is caused by amplification of the quantization noise for higher spatial frequencies in the reconstruction process. In the real reconstruction process, the quantization noise is accompanied by the computational noise resulting from precision and complexity of computation. Since the quantization noise is usually much higher than the computational one, it should have dominant effect on the propagation-distance range of image-reconstruction ability.

### 4. Conclusions

Existence of an inverse propagation operator for the Helmholtz-Kirchhoff propagation formula applied to the non-coherent wave field has been proved and its form has been found. However, the analysis of image reconstruction by means of this operator, realized in numerical way, shows existence of the propagation-distance limit, at which SNR in the reconstructed image rapidly decreases. This limit depends on the greatest spatial frequency occurring in the original image and on the number of bits of ADC and computing precision.

### Acknowledgements

This work is supported by the State core funding for statutory R & D activities of the Institute of Physics, University of Rzeszow, and Department of Physics, Rzeszow University of Technology, both channeled entirely through the Polish Ministry of Science and Higher Education